\documentclass[apj]{emulateapj}
\usepackage[colorlinks]{hyperref}
\hypersetup{ colorlinks, linkcolor=red, citecolor=blue }

\begin{document}

\title{Predicting the amount of hydrogen stripped by the supernova explosion for SN~2002cx-like SNe Ia}

\author{Zheng-Wei. Liu\altaffilmark{1,2,3,4}, M. Kromer\altaffilmark{4}, M. Fink\altaffilmark{6}, R. Pakmor\altaffilmark{5}, F. K. R\"opke\altaffilmark{6}, \\X. F. Chen\altaffilmark{1,2}, B. Wang\altaffilmark{1,2} and Z. W. Han\altaffilmark{1,2}}
\email{Email: zwliu@ynao.ac.cn} 

\altaffiltext{1}{Yunnan Observatories, Chinese Academy of Sciences, Kunming 650011, China}
\altaffiltext{2}{Key Laboratory for the Structure and Evolution of Celestial Objects, Chinese Academy of Sciences, Kunming 650011, China}
\altaffiltext{3}{University of Chinese Academy of Sciences, Beijing 100049, China}
\altaffiltext{4}{Max-Planck-Institut f\"ur Astrophysik, Karl-Schwarzschild-Str. 1, 85741 Garching, Germany}
\altaffiltext{5}{Heidelberger Institut f\"ur Theoretische Studien, Schloss-Wolfsbrunnenweg 35, 69118 Heidelberg, Germany}
\altaffiltext{6}{Institut f\"ur Theoretische Physik und Astrophysik, Universit\"at W\"urzburg, Am Hubland, 97074 W\"urzburg, Germany}

\begin{abstract}
 
The most favored progenitor scenarios for Type Ia supernovae (SNe Ia) involve the 
single-degenerate (SD) scenario and the double-degenerate scenario. The absence of 
stripped hydrogen (H) in the nebular spectra of SNe Ia 
challenges the SD progenitor models. Recently, it was shown that pure deflagration 
explosion models of Chandrasekhar-mass white dwarfs ignited off-center reproduce the characteristic 
observational features of 2002cx-like SNe Ia very well. In this work we predict, for the first time,
the amount of stripped H for the off-center pure deflagration explosions. We find that 
their low kinetic energies lead to inefficient H mass stripping ($\lesssim0.01\,\rm{M_{\odot}}$), indicating that the
stripped H may be hidden in (observed) late-time spectra of SN 2002cx-like SNe Ia.   
    
\end{abstract}

\keywords{supernovae: general --- binaries: close --- methods: numerical}

\section{INTRODUCTION}
 \label{sec:introduction}

   Type Ia supernovae (SNe Ia) are instrumental as distance indicators  on a
   cosmic scale to determine the expansion history of the Universe
   \citep{Ries98, Schm98, Perl99}. They are widely believed to 
   be caused by thermonuclear explosions of carbon/oxygen white 
   dwarfs (C/O WDs) in binary systems. 
   The two favored classes of SN Ia progenitors are the single-degenerate (SD) scenario
   and double-degenerate (DD) scenario. In the DDS, two C/O WDs merge due to gravitational
   wave radiation, leading to a SN Ia thermonuclear explosion (DD scenario, e.g., \citealt{Iben84}). 
   In the SD scenario, WDs accrete H/He-rich matters from companions that could be 
   main-sequence (MS) stars, sub-giants, red giants (RGs) or He stars. They  
   ignite SN Ia explosions when approaching the Chandrasekhar-mass ($M_{\rm{Ch}}$) 
   limit (e.g., \citealt{Whel73,Hach96, Han04}).

   Recently, some observational and hydrodynamical studies (see, e.g., \citealt{Li11, Nuge11, 
   Chom12, Hore12, Bloo12, Scha12, Pakm10, Pakm11, Pakm12b}) support the viability of DD scenario. 
   There are some observational indications (see, e.g., \citealt{Pata07, 
   Ster11, Fole12, Dild12}), suggesting that the  progenitors of some  SNe Ia 
   may come from the SD scenario. However, the exact nature of 
   SN Ia progenitors remains uncertain (see \citealt{Hill00, Hill13} for reviews).

    The spectra of normal SNe Ia are characterized by the absence of H and He lines 
    and a strong silicon absorption feature. To date, no direct observation shows the 
    signature of H lines in late-time, nebular spectra of SNe Ia \citep{Matt05, Leon07, Shap13}. 
    One of the signatures of the SD scenario is that the SN Ia explosion is expected 
    to remove H/He-rich material from its non-degenerate companion star \citep{Whee75}. 
    Hydrodynamical simulations with a classical SN Ia explosion model (i.e., the 
    W7 model, see \citealt{Nomo84}) showed that about $0.1\,\mathrm{M}_{\odot}$ 
    H-rich material are expected to be stripped off from a MS companion star by the impact 
    of the SN~Ia ejecta (see, e.g., \citealt{Mari00, Pakm08, Liu12, Liu13, Pan12}). 
    Almost the whole envelope of a RG companion ($\sim0.5\,\mathrm{M}_{\odot}$) is 
    removed (see, e.g., \citealt{Mari00, Pan12}). The amount of stripped H is significantly 
    above the most stringent upper limits on non-detection of H ($\sim\,$0.01--0.03$\,\mathrm{M}_{\odot}$, see \citealt{Leon07, Lund13}) 
    which were derived from observations of normal SNe~Ia. Moreover, \citet{Shap13} obtained a lower limit
    of $\sim0.001\,\mathrm{M}_{\odot}$ on detection of stripped H for SN 2011fe which is the
    nearest SN Ia in the last 25 years and has been observed in unprecedented detail. 
    Therefore, the absence of $\rm{H_{\alpha}}$ in late
    time nebular spectra of SNe Ia poses some problems for the SD scenario and favors other progenitor channel such as 
    WD merger (see \citealt{Pakm10, Pakm11, Pakm12b}). However, all previous 
    hydrodynamical simulations were performed with the classical W7 explosion model which
	is suitable for normal SNe Ia in nickel production and kinetic energy.

\begin{figure}
  \begin{center}
    {\includegraphics[width=0.48\textwidth, angle=360]{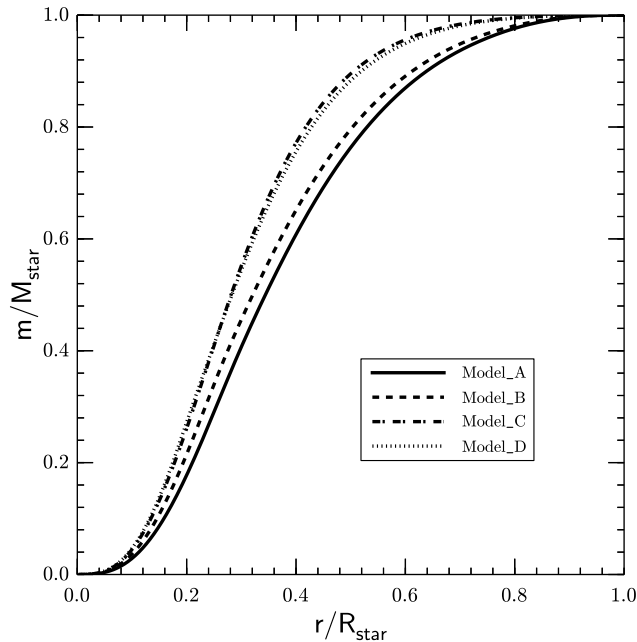}}
 \caption{Mass vs. radius profiles for four main-sequence companion models used in 
          the impact simulations.}
\label{Fig:1}
  \end{center}
\end{figure}

    SN 2002cx-like SNe are spectroscopically peculiar and faint objects compared to other
    SNe Ia. Their spectra are characterized by very low expansion velocities and show strong mixing of 
    the explosion ejecta \citep{Jha06, Phil07}. Moreover, SN 2002cx-like SNe are proposed to 
    originate from Chandrasekhar mass deflagrations, i.e., SD H-accreting progenitors.  
    Very recently, \citet{Krom13} performed hydrodynamics (see also \citealt{Jord12}) and radiative-transfer 
    calculations for a three-dimensional (3D) full-star pure deflagration 
    model (i.e., the N5def model, see \citealt{Fink13})  which is able to reproduce 
    the characteristic observational features of SN 2005hk (a prototypical 2002cx-like SN Ia). In 
    the N5def model, only a part of the $M_{\rm{Ch}}$ WD, 
    $\sim0.37\,\rm{M_{\odot}}$, is ejected with a  much lower kinetic energy 
    ($\sim\mathrm{1.34\times10^{50}\,erg}$) than models for normal SNe Ia. The thermonuclear explosion fails 
    to completely unbind the WD and leaves behind a bound remnant of $\sim1.03\,\rm{M_{\odot}}$
    which consists mainly of unburned C/O (see \citealt{Jord12, Krom13, Fink13}).     
    The small amount of kinetic energy released in this pure deflagration 
    model might significantly decrease the stripped companion mass,
    potentially avoiding a signature of H lines in late-time
    spectra of SN 2002cx-like SNe Ia.

    Here we calculate the amount of stripped H for the N5def model (which is the best current model
	for SN 2002cx-like SNe~Ia) in the SD scenario using 3D hydrodynamical simulations of 
	the impact of the SN ejecta on MS companion stars.
    The paper is organized as follows. In Section~\ref{sec:code}, we describe the methods and codes used 
    in this work. Section~\ref{sec:simulations} presents the results from hydrodynamical simulations. 
    The distribution of the unbound mass from population synthesis calculations is shown in Section~\ref{sec:bps}. 
    Some discusions based on results of impact simulations are presented in Section~\ref{sec:dis}.
	Finally, we summarize the basic results of simulations in Section~\ref{sec:summary}.

\begin{figure}
  \begin{center}
    {\includegraphics[width=0.48\textwidth, angle=360]{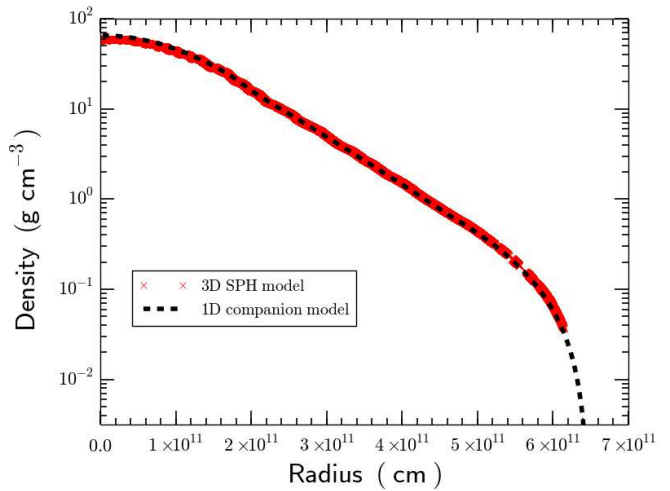}}
 \caption{Radial profiles of the density for a 1D companion model (i.e., Model\_A) and its corresponding
          3D SPH model generated by the healpix method (see \citealt{Pakm12a}).}
\label{Fig:2}
  \end{center}
\end{figure}

\begin{table*} \renewcommand{\arraystretch}{1.5}
\begin{center}
\caption{Initial models and results from hydrodynamical simulations. \label{table:1}}
\begin{tabular}{lccccccccc}
\hline\hline
Model & $M_{\rm{WD}}$ & $M_{\rm{2,i}}$ & $M_{\rm{2,f}}$ & $R_{\rm{2,f}}$ & $a_{\rm{f}}$&$\Delta M^{\mathrm{Def}}$ &$\Delta M^{\mathrm{W7}}$ & $v_{\mathrm{kick}}^{\rm{Def}}$ & $v_{\mathrm{kick}}^{\rm{W7}}$ \\
&$(\mathrm{M}_{\odot})$ & $(\mathrm{M}_{\odot})$ & $(\mathrm{M}_{\odot})$ &\multicolumn{2}{c}{$(\mathrm{10^{11}\,cm})$}&\multicolumn{2}{c}{$(\mathrm{M}_{\odot})$} &\multicolumn{2}{c}{ $(\mathrm{km\,s^{-1}})$} \\ 
\hline
Model\_A & 0.8 & 2.2 &1.21 & 0.65  & 1.77 & 0.015 & 0.173 &24.4 & 105.3\\
Model\_B & 0.9 & 2.4 &1.40 & 0.74  & 1.95 & 0.016 & 0.172 &22.8 & 94.9\\
Model\_C & 1.0 & 2.4 &1.88 & 0.91  & 2.25 & 0.013 & 0.116 &15.9 & 66.9\\
Model\_D & 1.2 & 3.0 &2.45 & 1.13  & 2.64 & 0.016 & 0.141 &14.9 & 65.3\\

\hline
\end{tabular}
\tablecomments{\footnotesize $M_{\rm{WD}}$ and $M_{\rm{2,i}}$ present the WD and companion mass
 at the beginning of mass transfer. $M_{\mathrm{2,f}}$, $R_{\rm{2,f}}$ and $a_{\rm{f}}$ demonstrate the companion mass, companion radius and 
 binary separation at the time of the explosion. $\Delta M^{\mathrm{Def}}$ 
 and $\Delta M^{\mathrm{W7}}$ show the unbound companion masses
 in the impact simulations for the pure deflagration model and the W7 model. $v_{\mathrm{kick}}^{\rm{\rm{W7}}}$
 and $v_{\mathrm{kick}}^{\rm{Def}}$ correspond to the companion kick velocities.  }
\end{center}
\end{table*}

\section{Numerical method and model}
\label{sec:code}

    In order to construct a detailed companion structure at the moment of the SN Ia explosion, we used 
    the same method as described in \citet{Liu12} to trace binary evolution in which a WD accretes 
    H-rich material from a MS companion star (i.e., WD+MS $M_{\rm{Ch}}$ explosion scenario). We think the WD 
    would explode as a SN Ia when its mass increases to the $M_{\rm{Ch}}$ limit. Here, we adopted the Eggleton's 
    stellar evolution code, Roche-lobe overflow and the optically thick wind model 
    of \citet{Hach96} were included into the code to treat mass transfer in the binary. 
    With a series of consistent binary evolution calculations, we selected four companion star models as 
    input models of hydrodynamical simulations. These four companion stars were constructed with different 
    initial WD masses, companion masses and orbital periods (see Table~\ref{table:1}), which leads to 
    companion models different in mass, orbital period and detailed structure at the moment of the 
    SN explosion. Four companion models created in 1D binary evolution calculations are summarized 
    in Table~\ref{table:1}, their radial mass profiles 
    are shown in Figure~\ref{Fig:1}. We then performed 3D hydrodynamical simulations of the impact of SN Ia ejecta on 
    the companion star employing the SPH code {\sc Stellar~GADGET} \citep{Pakm12a, Spri05}.  
    
    In this work, all initial conditions and basic setup for the impact 
    simulations are the same as those in \citet{Liu12}. 
    We use the healpix method described in \citet{Pakm12a} to map the 1D profiles of density and
    internal energy of a 1D companion star model to a particle
    distribution suitable for the SPH code.  Before we start the
    actual impact simulations, the SPH model of each companion star is 
    relaxed for several dynamical timescales to reduce numerical noise introduced 
    by the mapping. A comparison of density profiles between the 1D stellar model
    and its consistent SPH model for Model\_A are shown in Figure~\ref{Fig:2}.

    The SN  Ia explosion was represented by the pure deflagration model 
    of \citet{Krom13} (i.e., the N5def model). This model has been shown 
    to reproduce the characteristic observational features of 
    2002cx-like SNe Ia well \citep{Krom13, Jord12}. In this simulation,  only the $0.37\,\rm{M_{\odot}}$ of 
    ejected material with a total kinetic energy of $\rm{1.34\times10^{50}\,erg}$ were 
    used to represent the SN Ia explosion, we did not include the $1.03\,\rm{M_{\odot}}$ 
    bound remnant of the $M_{\rm{Ch}}$ WD into the simulations.
    Based on the angle averaged 1D ejecta structure of the N5def model, 
    SPH particles were placed randomly in shells to reproduce the mass 
    (density) profile and gain the radial velocities they should have 
    at their positions. The composition of a particle was then set to 
    the values of the initial 1D model at a radius equal to the radial 
    coordinate of the particle. Here, the effect a mild degree of asymmetry caused
    by off-center pure deflagration explosion of a $M_{\rm{Ch}}$ WD
    was ignored. However, only in the direction opposites 
    to the one-sided ignition region the velocities are somewhat lower \citep{Fink13}.
    The orientation of an asymmetry of SN ejecta plays an inefficient role in mass stripping
    by the time of interaction with the companion.\footnote[1]{We expect 
    that somewhat lower velocity (lower kinetic energies of SN ejecta) in the direction opposites to the 
    one-sided ignition region would lead to a little smaller total stripped mass.}

    We used $6\times10^{6}$ million SPH particles 
    to represent the He companion stars
    in all simulations of this work.\footnote[2]{Our previous work have concluded that it is 
    sufficient to represent the companion stars with about 6 million SPH particles 
    in the impact simulations to study the amount of unbound companion mass caused by
    the SN impact (see \citealt{Liu12}).} Because all SPH particle was set up with the same mass,
    the number of particles representing the supernova explosion is then fixed. 
    The supernova was placed at a distance to the companion star given by 
    the separation at the moment of SN Ia explosion in our 1D binary-evolution 
    calculations. The impact of the SN Ia ejecta on their binary companions 
    was then simulated for 5000 s, at which point the mass stripped off
    from the companion star and its kick velocity due to the impact have 
    reached constant values.

\begin{figure*}
  \begin{center}
    {\includegraphics[width=0.5\textwidth, angle=270]{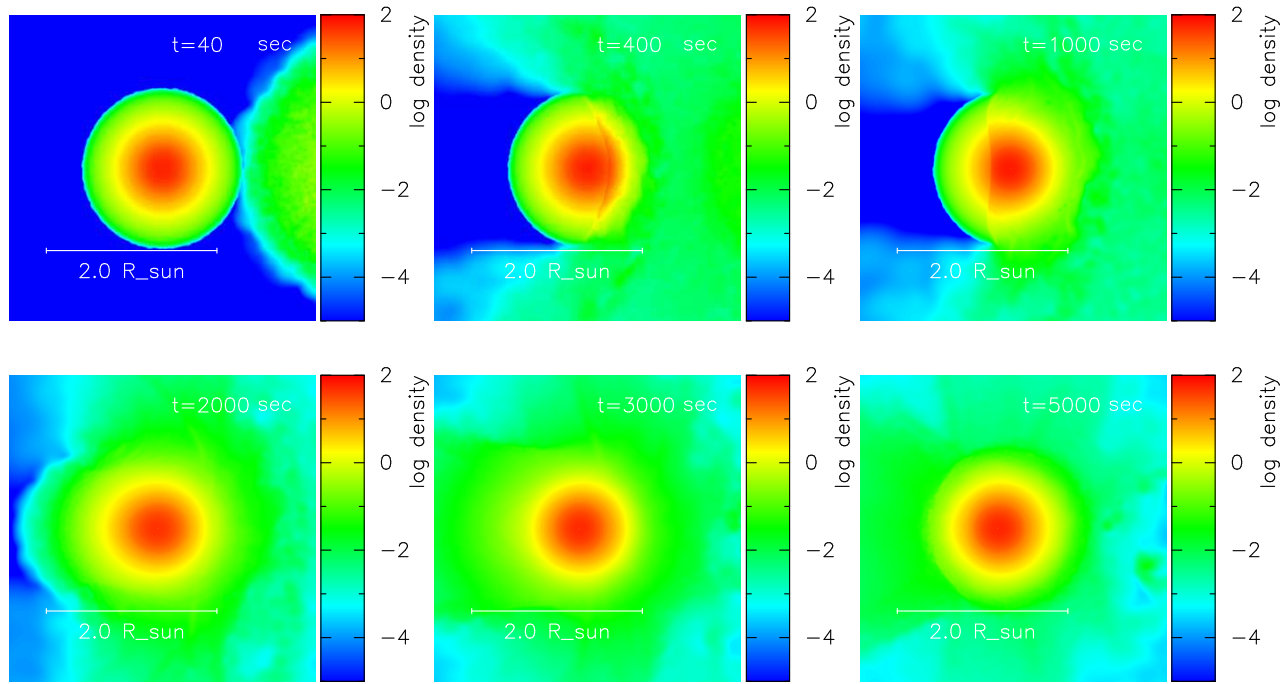}}
 \caption{Density distributions of all gas material as a function of the explosion time in the impact 
          simulations for Model\_A. The color scale shows the logarithm of the density in $\rm{g\,cm^{-3}}$. The plots are made with
          the freely available {\sc SPLASH} tool \citep{Pric07}.}
\label{Fig:rho}
  \end{center}
\end{figure*}

\section{Hydrodynamical results}
\label{sec:simulations}

 Figure~\ref{Fig:rho} shows the typical evolution of the density distribution in 
 the impact simulations for the companion star Model\_A. Compared to our previous hydrodynamical
 simulations for the same MS companion star model (but with a different explosion model,
 see \citealt{Liu12}), the basic impact processes are quite similar. The SN explodes at the right 
 side of the companion star. After the SN explosion, the SN ejecta expand freely for a while 
 and hit the MS companion star, removing solar-metallicity companion material and forming a 
 bow shock. Subsequently, the bow shock propagates through the companion star, causing an additional loss of 
 H-rich companion material from the far side of the star. Finally, the final unbound H-rich material of 
 the companion star caused by the SN impact is largely embedded in low-velocity
 SN debris behind the companion star, and the strongly impacted companion starts to relax to become almost
 spherical again.

\begin{figure}
  \begin{center}
    {\includegraphics[width=0.45\textwidth, angle=360]{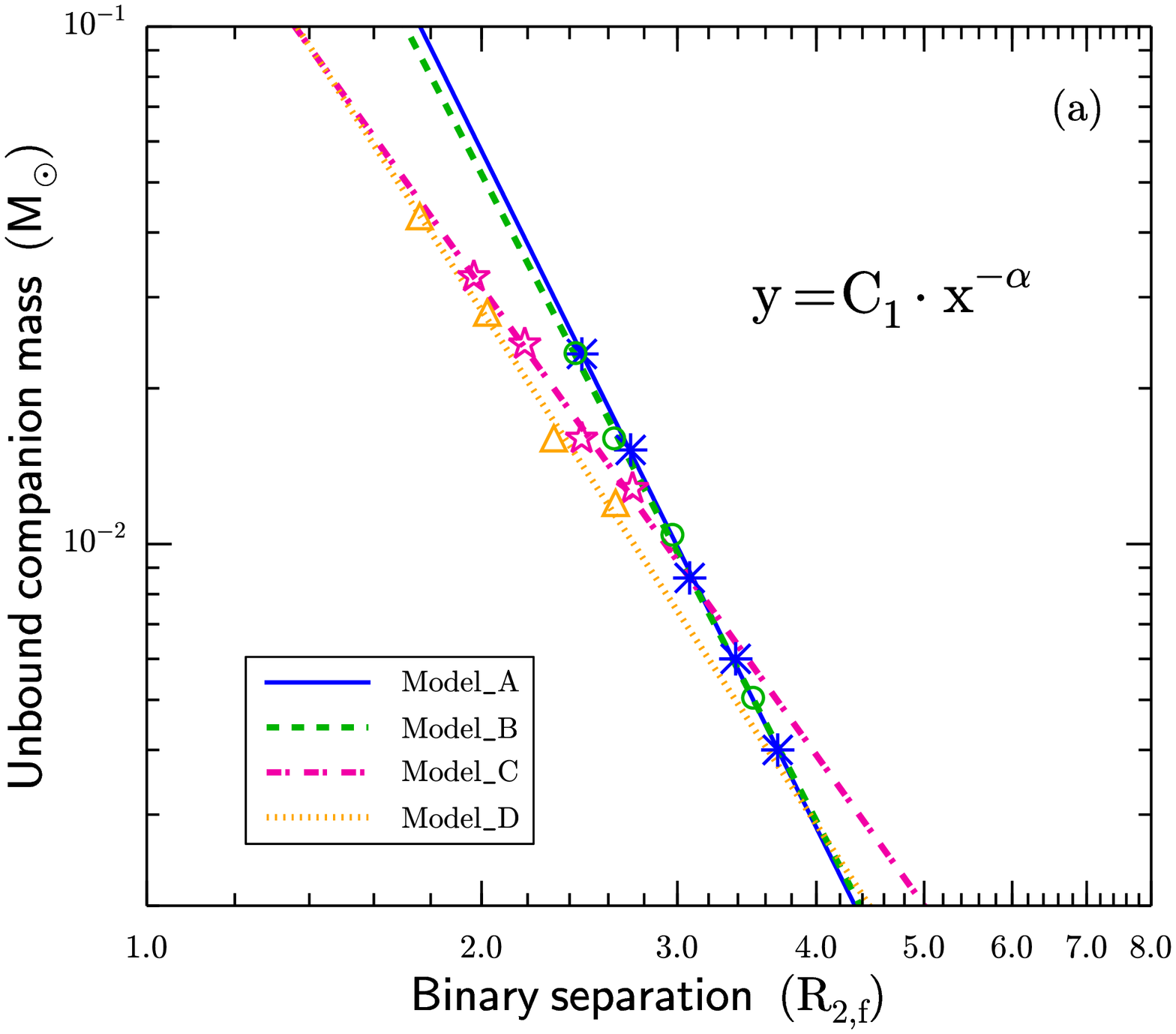}}
    {\includegraphics[width=0.45\textwidth, angle=360]{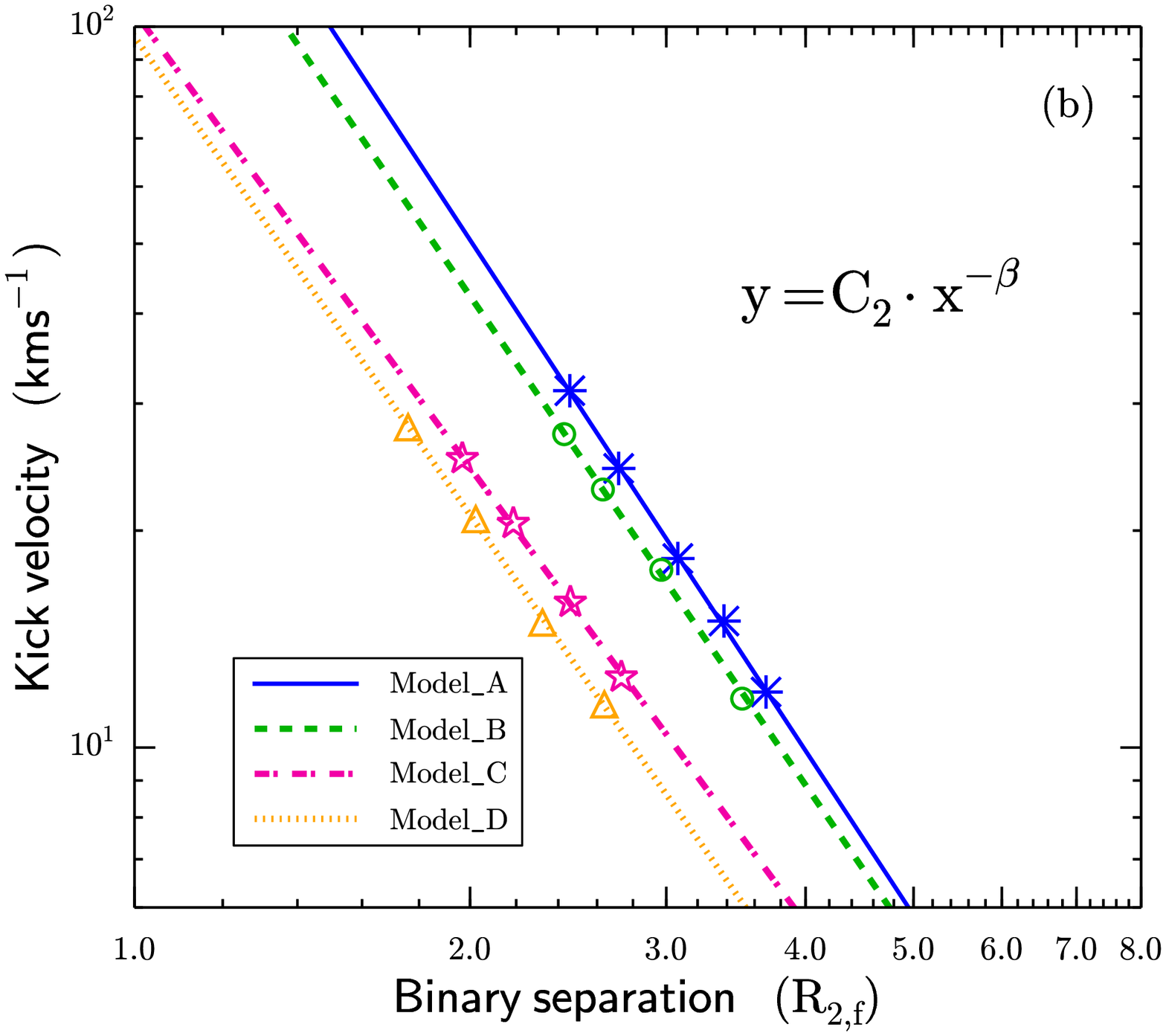}}
 \caption{Dependence of unbound companion mass ({\it panel a\/}) or kick 
          velocity ({\it panel b\/}) on the binary separation 
          in the impact simulations of four different companion models. The corresponding fitting parameters 
          are shown in Table~\ref{table:2}.} 
\label{Fig:fit}
  \end{center}
\end{figure}

After the SN explosion, the non-degenerate companion star is significantly hit
 by the SN ejecta. The SN impact remove H-rich material from outer layers of the 
 companion star through the ablation (SN heating) and the stripping (momentum transfer)
 mechanism. The late-time spectra of SN Ia probably show a signature of H lines 
 if a large amount of H-rich material can be stripped from the companion star during the 
 interaction with the SN ejecta.
 To calculate the amount of final unbound companion mass due to the SN 
 impact, we sum the masses of all unbound SPH particles (ablated+stripped 
 particles) that originally belongs to the 
 companion star at the end of the simulations.

For four different MS companion star models, our impact simulations show 
 that the companion star received a kick velocity of $\sim\rm{15}$ (Model\_D)--$\rm{25\,km\,s^{-1}}$ (Model\_A) at 
 the end of the simulations. Moreover, only a small amount of $\rm{0.013}$ 
 (Model\_C)--$\rm{0.016\,M_{\odot}}$ (Model\_D)
 of H-rich material is removed (ablation+stripping) from the companion stars 
 due to the SN impact (see Table~\ref{table:1}).

For a comparison, numerical 
 results for a classical explosion model  
 of a $M_{\rm{Ch}}$ WD for normal SNe Ia, the W7 model, are also shown in Table~\ref{table:1}.
 Because the N5def model does not burn the complete WD but leaves behind 
 a $\sim1.0\,\rm{M_{\odot}}$ bound remnant, it produces a much lower kinetic energy  
 ($\rm{1.34\times10^{50}\,erg}$) than the W7 model ($\rm{1.23\times10^{51}\,erg}$). Therefore, 
 a much smaller amount of H-rich material of only $\sim0.015\,\rm{M_{\odot}}$ is removed from 
 the companion stars by the impact of the pure deflagration explosion of a $M_{\rm{Ch}}$ WD. 
 In contrast, the amount of final unbound companion mass is more than ten times larger 
 ($>0.1\,\rm{M_{\odot}}$) for the W7 explosion model with the 
 same companion star models at the same separations.

A large amount of unbound companion mass in the impact simulations for the W7 explosion 
 model seems to indicate that normal SNe Ia are not likely produced from the WD+MS $M_{\rm{Ch}}$ 
 explosion scenario (see also \citealt{Liu12}). Here, it is shown that 
 the mass stripping is inefficient for
 2002cx-like SNe Ia due to the low kinetic 
 energies of off-center pure deflagration explosions of a $M_{\rm{Ch}}$ WD, leading to that
 a amount of unbound companion material 
 caused by the SN impact is quite small. 
 Therefore, the H lines probably be hidden in late-time 
 spectra of SN 2002cx-like SNe Ia. However, whether or not such small amount of stripped H mass would
 expect to show a signature of 
 H lines in late-time spectra of 
 SN 2002cx-like SNe Ia, which needs to analyze late-time 
 spectra of 2002cx-like SNe to obtain the lower limit of stripped mass for detecting the H 
 lines.\footnote[3]{Observationally, \citet{Leon07} obtained the 
 limit for detecting stripped H in nebular spectra of normal SNe Ia is $\rm{0.01\,M_{\odot}}$ (see also \citealt{Lund13}).}

\section{Population synthesis results}
\label{sec:bps}

\subsection{Power-law fitting}
Different WD+MS binary systems evolve to different evolutionary stages and have different 
binary parameters when the WD explodes as an SN Ia. Consequently, the companion radius 
and binary separation of binary systems differ significantly from the MS companion models used in our 
present simulations. We therefore investigated the dependence of the numerical results on the 
ratio of binary separation to the companion radius ($a_{\rm{f}}/R_{\rm{2,f}}$) at the time of the explosion. Here,
we used the same method as in \citet{Liu12} to artificially adjust the binary separations for
a fixed companion star model (which means that all parameters but the orbital separation
are kept constant).

\begin{table} \renewcommand{\arraystretch}{1.5}
\begin{center}
  \caption{Fitting parameters for equation~(\ref{equation:1})
    and~(\ref{equation:2}) \label{table:2}}
\begin{tabular}{lcccc}
\tableline\tableline
  & \multicolumn{4}{c}{Fitting parameters} \\
Model & $\mathrm{C_{1}}$ & $\mathrm{\alpha}$ &  $\mathrm{C_{2}}$ & $\mathrm{\beta}$\\ 
\tableline
Model\_A & 1.169 & 4.345 & 258.7  & 2.354 \\
model\_B & 0.251 & 3.003 & 201.9  & 2.254 \\
Model\_C & 0.923 & 4.153 & 104.8  & 2.100 \\
model\_D & 0.279 & 3.300 & 96.8   & 2.203 \\
\tableline
\end{tabular}
\end{center}
\end{table}

\begin{figure}
  \begin{center}
    {\includegraphics[width=0.5\textwidth, angle=360]{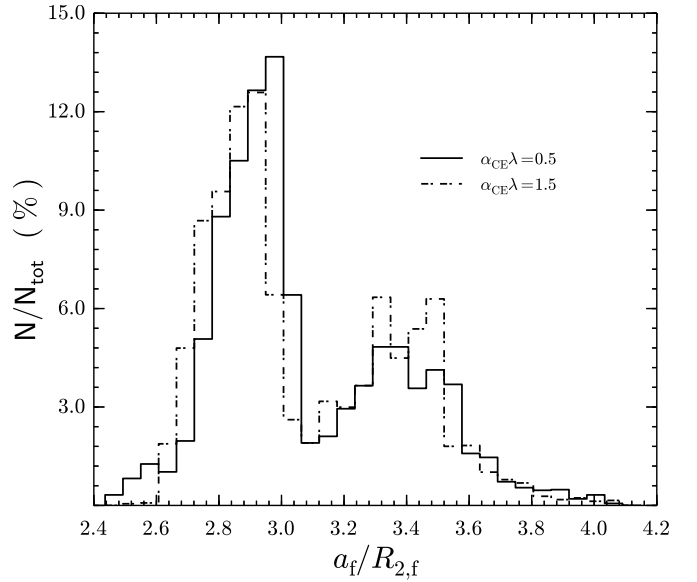}}
%    {\includegraphics[width=0.45\textwidth, angle=360]{f3b.eps}}
 \caption{Distribution of the ratio of binary separation to companion
          radius ($a_{\rm{f}}/R_{\rm{2,f}}$) in the population synthesis calculations for WD+MS progenitor models.
          The solid and dash-dotted line show results of the models with $\rm{\alpha_{CE}\lambda}=0.5$
          and $\rm{\alpha_{CE}\lambda}=1.5$ in WLH10.}
\label{Fig:mass}
  \end{center}
\end{figure}

Figure~\ref{Fig:fit} presents the amount of final unbound companion mass and kick velocity 
as a function of the parameter of $a_{\rm{f}}/R_{\rm{2,f}}$ for four companion star models. It
shows that the unbound mass and kick velocity significantly decrease when increasing the orbital 
separation of the binary. Generally, these relations can be fitted with power law functions 
in good approximation (see Figure~\ref{Fig:fit}):

  \begin{equation}
    \label{equation:1}
     M_{\mathrm{unbound}}= C_1  \left(\frac{a_{\rm{f}}}{R_{\rm{2,f}}}\right)^{-\alpha} \ \ \mathrm{M}_{\odot},
   \end{equation}

  \begin{equation}
    \label{equation:2}
     v_{\mathrm{kick}}= C_2  \left(\frac{a_{\rm{f}}}{R_{\rm{2,f}}}\right)^{-\beta} \ \ \mathrm{km\,s^{-1}},
   \end{equation}

   where $a_{\rm{f}}$ is the binary separation, $R_{\rm{2,f}}$ is the radius of the MS
   companion star at the onset of the SN explosion. $C_1$ and $C_2$ are two
   constant, $\alpha$ and $\beta$ are the power-law indices (see
   Table~\ref{table:2}).

\subsection{Unbound masses and kick velocities}
\label{sec:bps2}
 
\citet{Wang10b}(hereafter WLH10) performed comprehensive binary population 
synthesis (BPS) calculations obtaining a large sample of WD+MS SN Ia progenitor models. 
They predicted many properties of the companion stars and binary systems at 
the moment of the SN explosion (e.g. the companion masses, the companion radii, 
the orbital periods, etc). The distributions of the ratio of binary separations 
to the companion radii ($a_{\rm{f}}/R_{\rm{2,f}}$) in the WLH10 sample (their 
model with $\alpha_{\rm{CE}}\times\lambda=0.5, 1.5$) are presented in Figure~\ref{Fig:mass}. 
It shows that most systems are concentrated at $a_{\rm{f}}/R_{\rm{2,f}}\sim2.9$ and $3.4$.  
Based on the distribution of $a_{\rm{f}}/R_{\rm{2,f}}$, we calculated the final unbound companion masses (and kick velocities) 
due to the SN impact by adopting the power law relations of four companion star models in Figure~\ref{Fig:fit}, 
which are displayed in Figure~\ref{Fig:bps}. As it is shown, the impact of off-center pure deflagration 
explosions of $M_{\rm{Ch}}$ WDs lead to a small amount of mass loss of the companion 
star ($\lesssim0.015\,\rm{M_{\odot}}$ ) in almost 
all WD+MS binary systems from BPS calculations. Moreover,
the SN impact delivers a small kick velocity of $\lesssim25\,\rm{km\,s^{-1}}$ to
the companion star.

In Section~\ref{sec:code}, initial MS companion star models were set up 
with a H abundance of $\rm{X=0.7}$, a He abundance of $\rm{Y=0.28}$ and a metallicity 
of $\rm{Z=0.02}$ when we constructed the MS companion star model at the moment of the SN 
explosion. Therefore, pure deflagrations of the $M_{\rm{Ch}}$ WDs strip off a small amount of pure H of 
$\lesssim\mathrm{0.01\,M_{\odot}}$ from the companion stars in almost all WD+MS models.
These small stripped H masses are consistent with the lower mass limit for detecting $\rm{H_{\alpha}}$ 
emission lines in nebular spectra of normal SNe Ia ($\mathrm{0.01\,M_{\odot}}$, see \citealt{Leon07}). 
The inefficient mass stripping seems to imply the stripped H  may be hidden in (observed) 
late-time spectra of most of SN 2002cx-like SNe Ia.    
However, only normal SNe Ia were looked at when \citet{Leon07} obtained the upper limit for 
non-detection of stripped $\rm{H_{\alpha}}$ of $\mathrm{0.01\,M_{\odot}}$, the observational limits 
for SN 2002cx-like SNe may be smaller/larger than the value for normal SNe Ia. Most of stripped H-rich material in this 
simulations ends up at velocities below $10^{3}\,\rm{km\,s^{-1}}$ so that it 
is confined to the innermost part of explosion ejecta. Whether or not $\rm{H_{\alpha}}$ emission 
will be detectable, can only be answered by performing sophisticated radiative transfer simulations 
on the abundance structure of our explosion models.

\section{Discussion}
\label{sec:dis}

\subsection{The influence of the companion structures}

To test influence of the ratio of binary separation to the 
companion radius of $a_{\rm{f}}/R_{\rm{2,f}}$, we artificially 
adjusted the binary separation for a fixed companion star model.
The final unbound mass due to the SN impact
decreases by a factor of 10 as the parameter of $a_{\rm{f}}/R_{\rm{2,f}}$
increases by a factor of 2. 

Figure~\ref{Fig:fit} shows that  the fitting 
parameters of four different companion models are different, which indicates that the companion 
structure results from the characteristics of the original binary system
and the details of the mass transfer also can affect the final unbound mass.  
In reality, the companion structure is not independent on the binary
separation in the binary evolutions. Therefore, the companion structure would be 
different with different $a_{\rm{f}}/R_{\rm{2,f}}$ from BPS calculations.

\begin{figure*}
  \begin{center}
    {\includegraphics[width=0.45\textwidth, angle=360]{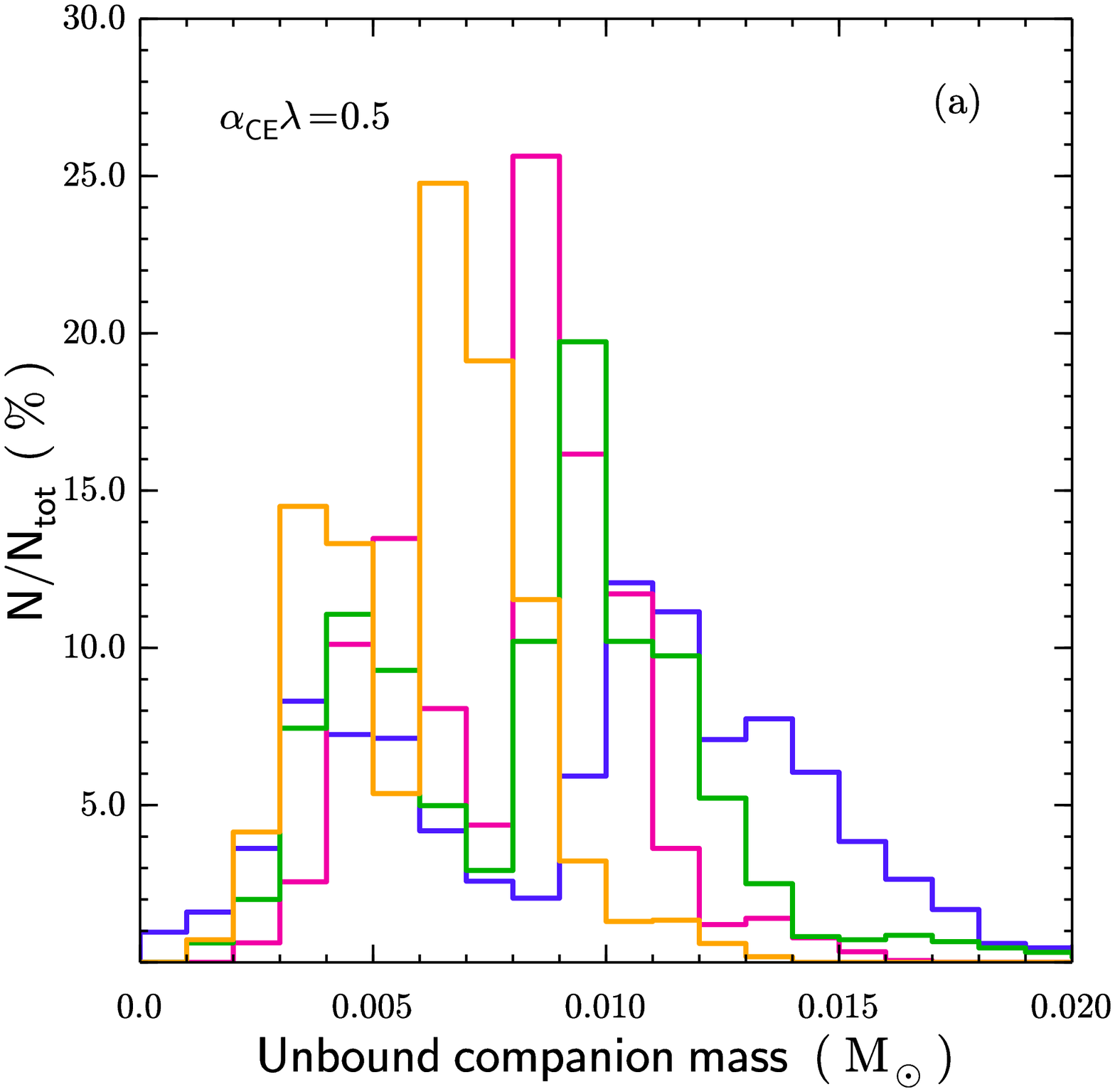}}
    {\includegraphics[width=0.45\textwidth, angle=360]{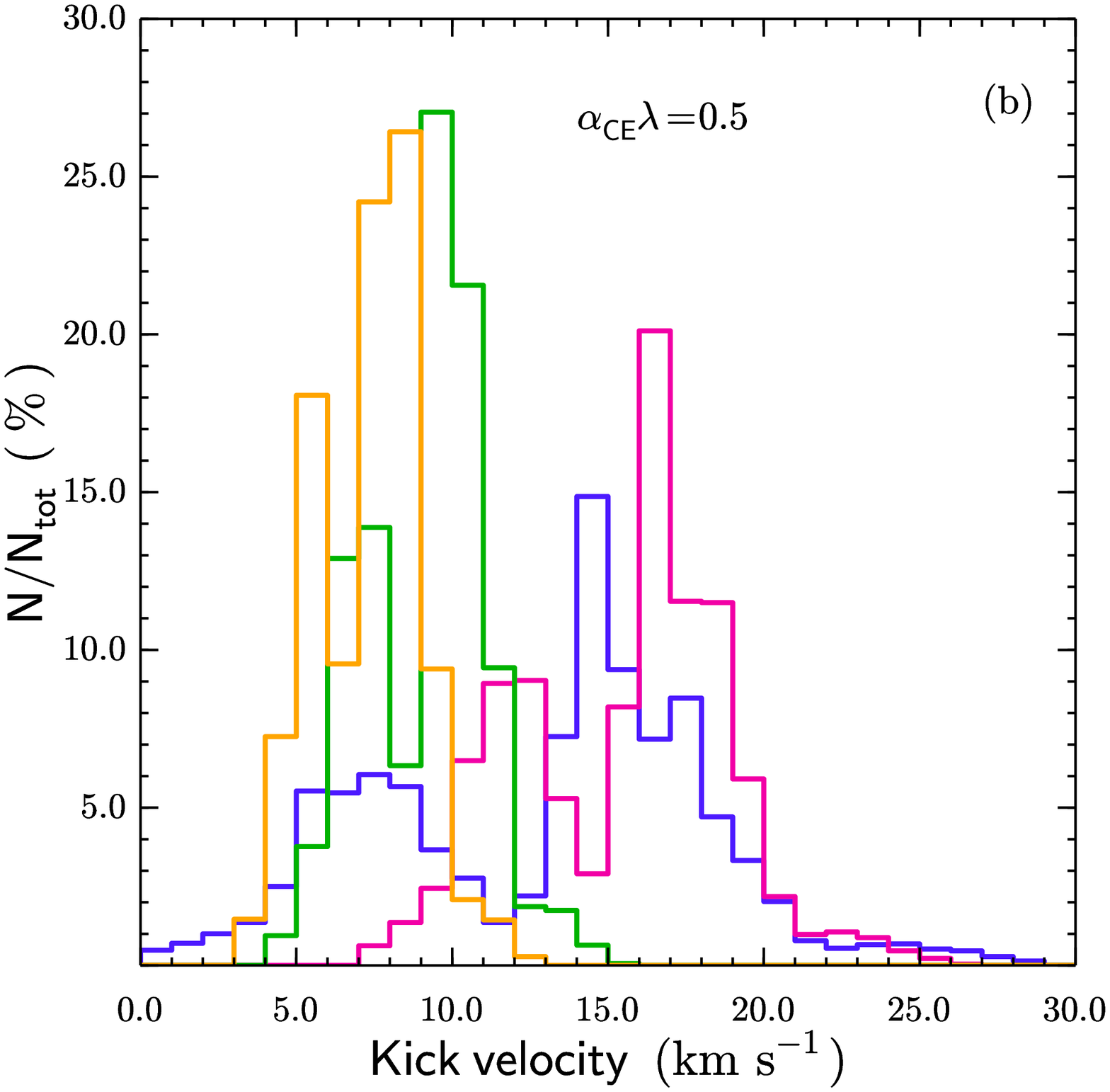}}
    {\includegraphics[width=0.45\textwidth, angle=360]{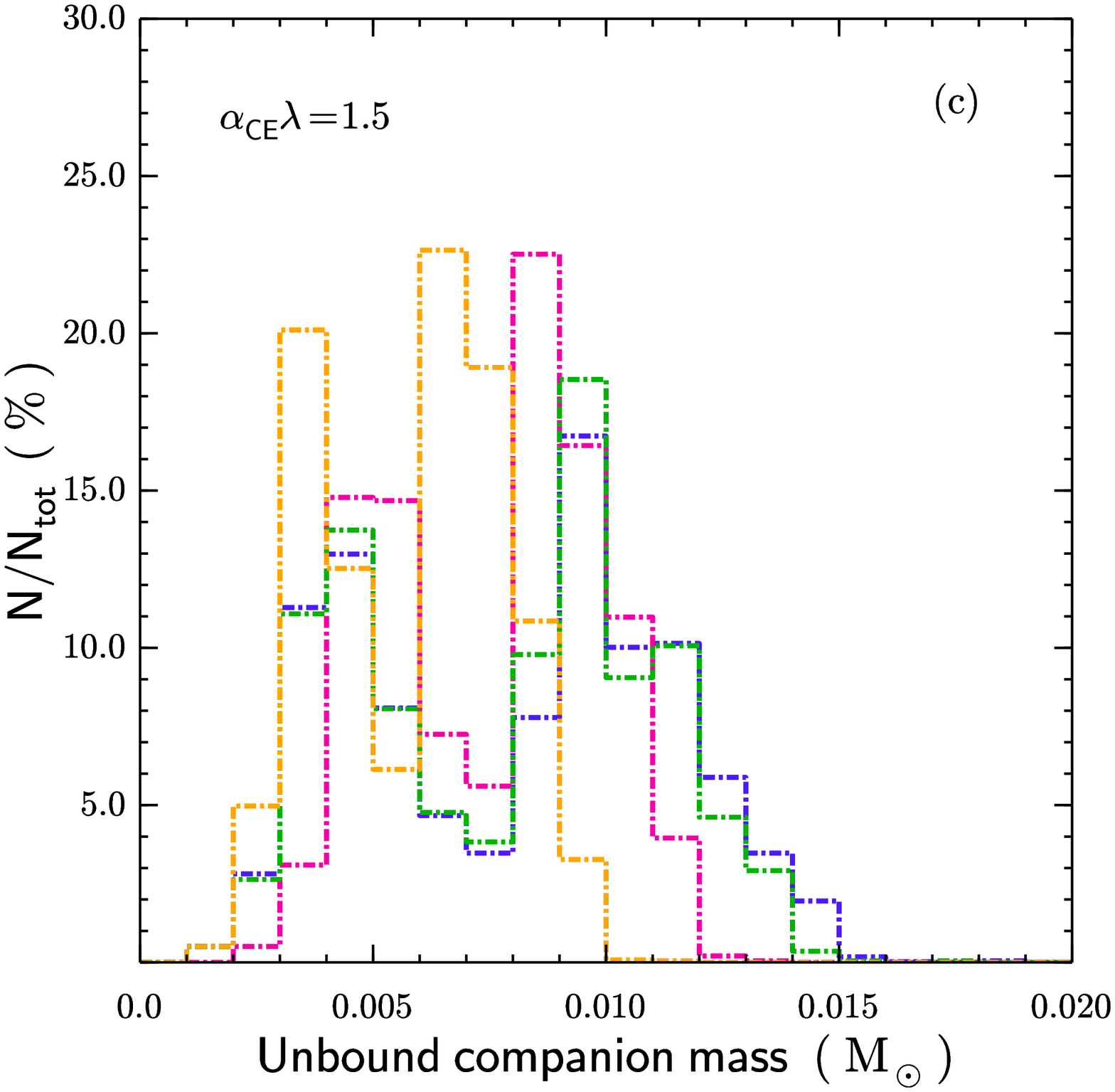}}
    {\includegraphics[width=0.45\textwidth, angle=360]{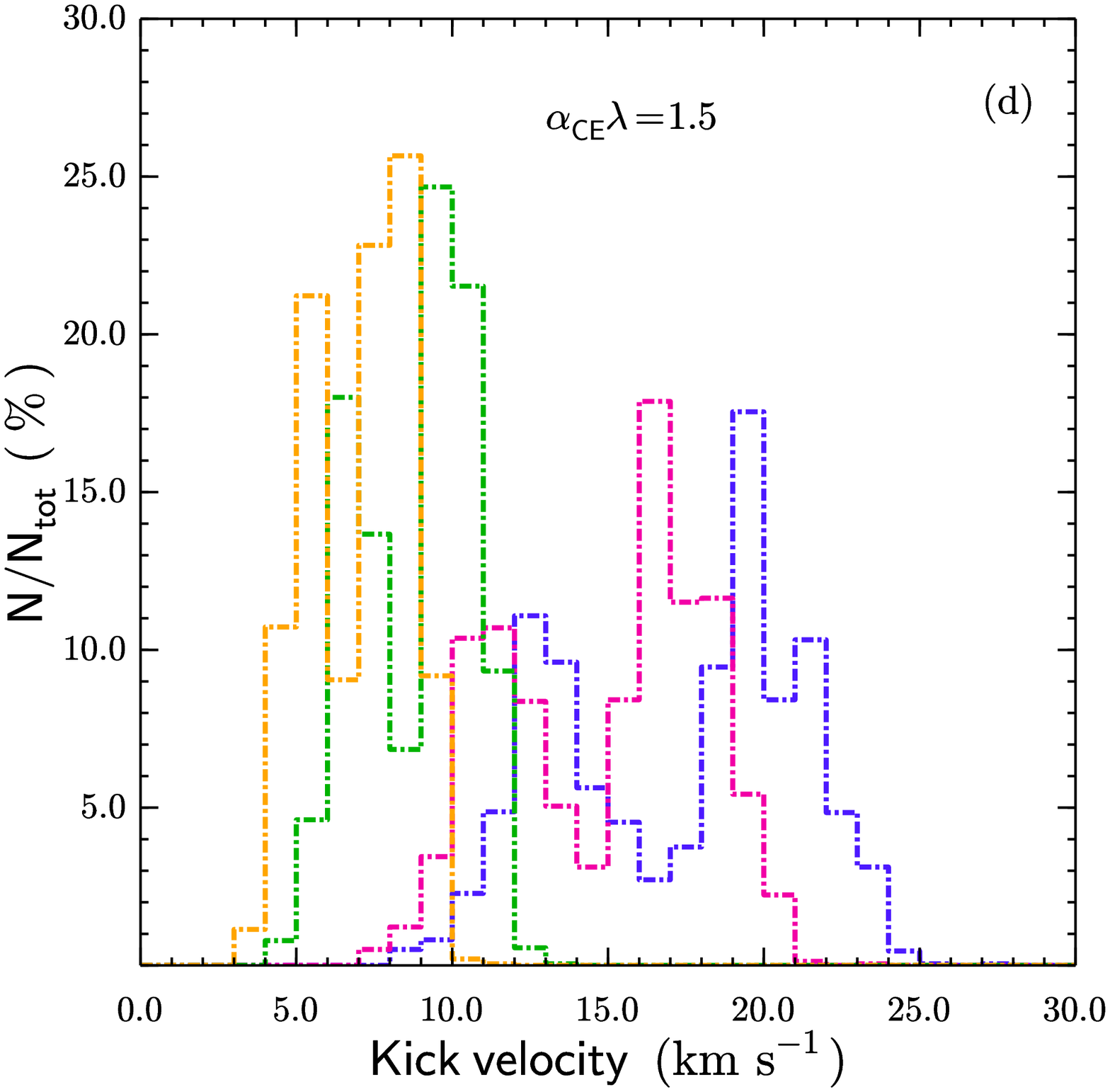}}
 \caption{  Distributions of the final unbound companion 
            mass (first cloumn) and the kick velocity (second cloumn) due to the SN impact 
            in the simulations. Different color show the results that are calculated by using 
            the relation obtained from the power-law fitting (see Figure~\ref{Fig:fit}) 
            for Model\_A (bule lines), Modle\_B (red lines), Model\_C (green lines), and Model\_D (yellow lines).
            The solid (top row) and dash-dotted lines (bottom row) show results of the models with $\rm{\alpha_{CE}\lambda}=0.5$
            and $\rm{\alpha_{CE}\lambda}=1.5$ in WLH10.}
\label{Fig:bps}
  \end{center}
\end{figure*}

In Section~\ref{sec:bps2}, we only used the same power-law relation obtained from a fixed companion star 
model (for example, the power-law relation in Model\_A) to calculate the total 
mass loss of the companion star caused by the SN impact for all WD+MS models, 
which ignores the influence of details of the companion 
structure. Figure~\ref{Fig:bps} show that a comparison of the results
that calculated by using four different power-law relations between the final unbound mass (and the kick velocities) and 
the parameter of  $a_{\rm{f}}/R_{\rm{2,f}}$ in Model\_A, Model\_B, Model\_C, and Model\_D.
Some differences (but not big differences) in the distribution of final unbound masses are 
seen in Figure~\ref{Fig:bps}a,c, which 
implies that the parameter of $a_{\rm{f}}/R_{\rm{2,f}}$ is not the only 
factor to determine the final stripped companion mass, the companion structure can also affect
the results.

\subsection{Explosion energy}

For a comparison, we performed the impact simulations for the same companion star 
by adopting both the N5def model and the W7 explosion model to represent the SN Ia explosion. A factor of 10 lower 
explosion energy in the N5def model leads to the stripped material reduced by a factor of 10 (see Table~\ref{table:1}). 
Moreover, \citet{Pakm08} investigated the influence of the SN explosion energy on 
the interaction with the companion star. They also found that explosion energy range 
covers a factor of 2 therefore leads to the unbound companion mass varies by a factor of 2.    
These results indicate that the SN explosion energy have only a small effect on the 
total mass loss of the companion star as compared to the effect of the parameter
of $a_{\rm{f}}/R_{\rm{2,f}}$ discussed above. Therefore, the ratio of the binary 
separation to the radius of companion star ($a_{\rm{f}}/R_{\rm{2,f}}$) is the most
important parameter to determine the final unbound companion mass (see also \citealt{Liu12, Liu13b}).

\subsection{The class of SN 2002cx-like SNe}

The SN 2002cx was discovered as a new class of peculiar SNe Ia by 
\citet{Li03}. From a volume-limited sample of the Lick Observatory Supernova Search (LOSS),
\citet{Li03} estimate that SN 2002cs-like SNe Ia contribute at about 5 per cent to the total
SN Ia rate. Very recently, \citet{Fole13} concluded that ``SNe Iax''  (the
prototype of which is SN 2002cx SNe) are the most common “peculiar” class of SNe,
they estimated that in a given volume SNe Iax could contribute $\sim1/3$ of total SNe Ia. Nonetheless, 
to date, only 25 SNe Iax are confirmed that they are observationally similar to its 
prototypical member, SN 2002cx (see \citealt{Fole13}). This
sample consists of 25 members is a very small fraction of total SNe Ia. In this work, the results obtained from the impact simulations only 
apply to the subclass of peculiar 2002cx-like SNe but not the bulk of SNe Ia. Therefore,
even the off-center pure deflagration explosion of a $M_{\rm{Ch}}$ WD removes 
H-rich material during the interaction with the MS companion star, 
the H stripped from the companion star may not be observed in the ejecta 
of such relatively rare events.

\subsection{Post-explosion fate of the binary}
\label{sec:fate}

In this work, the N5def pure deflagration model was used to represent the SN Ia explosion in our 
impact simulations. The hydrodynamics calculations of \citet{Jord12} and \citet{Krom13} showed that the N5def model does not
burn the complete WD but leaves behind a $\rm{\sim1.0\,M_{\odot}}$ bound remnant. Unfortunately, 
this bound remnant cannot be properly spatially resolved until late time in hydrodynamical 
simulations due to the strong expansion of the SN ejecta.

Our simulations show that the companion star receives a small kick 
velocity ($<30\,\rm{km\,s^{-1}}$ ) during the interaction with the SN ejecta. Therefore, 
whether the WD+MS binary system would be destoried after the SN explosion,  
which depends primarily on the kick velocity of the bound remnant.   
If the bound remnant receives
a large kick velocity that can overcome its gravitational force, the existence of abundance-enriched MS-like stars    
and WDs with peculiar spatial velocities are indicators (\citealt{Jord12}) of this studied progenitor scenario. Otherwise, 
the new binary system would survive the SN explosion.\footnote[4]{After the SN explosion, 
it was found that the bound remnant receives small kick velocities of $\sim36\,\rm{km\,s^{-1}}$
 (see \citealt{Krom13}) or large kick velocities up
to $\rm{520\,km\,s^{-1}}$ (see \citealt{Jord12}). The difference of kick velocity of 
bound remnant may originate from the different gravity solvers used.}  In this study, it is found that the post-impact radii of 
companions at $5000\,\rm{s}$ after the explosion are larger than the initial binary separations used for the impact simulations due to
their extreme expansions caused by the SN heating.\footnote[5]{However,
the effect of the $1.03\,\rm{M_{\odot}}$ bound remnant of the $M_{\rm{ch}}$ WD was not considered in our impact simulations.} 
This indicates that the surviving binary system may evolve and merge into a single object with a rapid rotation velocity, or 
experience a common envelope phase. The details of this post-explosion evolution should be 
addressed in future work. However, fully resolving the detailed structure of the bound remnants 
is a prerequisite for this investigation.

\section{Summary and conclusions}
\label{sec:summary}

We presented 3D hydrodynamical simulations of the impact of SN Ia explosions on their companion 
stars for the WD+MS scenario for the pure deflagration model presented in \citet{Krom13}. 
For four different companion star models, we find that the much lower kinetic energy of the
pure deflagration model compared to models for normal SNe Ia leads to a much lower stripped H-rich mass 
of only $\mathrm{0.013\,M_{\odot}}$--$\mathrm{0.016\,M_{\odot}}$.

Moreover, using the distribution of $a_{\rm{f}}/R_{\rm{2,f}}$ from BPS calculations, we 
discussed the distribution of the amount unbound H-rich material of the companion star using 
a power law relation between the total unbound mass and the ratio of binary separation to the 
companion radius ($a_{\rm{f}}/R_{\rm{2,f}}$). We find 
that the off-center pure deflagration explosions strip off a small amount 
of H ($\lesssim\mathrm{0.01\,M_{\odot}}$) from MS companion stars in most WD+MS progenitor 
models of SN 2002cx-like SNe~Ia. The inefficient mass stripping
may lead to the stripped H is fully hidden in their late-time 
spectra. Therefore, it will be very interesting to analyze late-time 
spectra of 2002cx-like SNe~Ia for the presence of hydrogen emission.

\section*{Acknowledgments}

%\acknowledgments

Z.W.L acknowledges the financial support from the MPG-CAS Joint Doctoral Promotion
Program (DPP) and from Max Planck Institute for Astrophysics (MPA).  This
work is supported by the National Basic Research Program of China
(Grant No. 2009CB824800), the National Natural Science Foundation of
China (Grant Nos. 11033008 and 11103072) and the Chinese Academy of
Sciences (Grant N0. KJCX2-YW-T24). The work of F.K.R was supported by 
Deutsche Forschungsgemeinschaft via the Emmy Noether Program (RO 3676/1-1) 
and by the ARCHES prize of the German Federal Ministry of Education 
and Research (BMBF). The simulations were carried out at the Computing 
Center of the Max Plank Society, Garching, Germany.

\
\
\
\
\
\ 

\
%\bibliographystyle{apj}

%\bibliography{ref}

\clearpage

%\appendix

%\section{Appendix material}

\end{document}